\documentclass[final,5p,times,twocolumn]{elsarticle}

\usepackage{amsmath}
\usepackage{slashed}
\usepackage{amssymb}
\usepackage[retainorgcmds]{IEEEtrantools}
\usepackage[colorlinks=true, linktoc=page, citecolor=blue, urlcolor=blue]{hyperref}
\usepackage{simpler-wick}
\usepackage{tikz-feynman}
\usepackage{float}

\newcommand{\N}{\mathcal{N}}
\newcommand{\K}{\mathcal{K}}
\newcommand{\W}{\mathcal{W}}
\newcommand{\LL}{\mathcal{L}}
\newcommand{\I}{\mathcal{I}}

\newcommand{\D}{\mathbb{D}}

\journal{Physics Letters B}

\hypersetup{pdfstartview={XYZ null null 0.75}}
\allowdisplaybreaks

\begin{document}

\begin{frontmatter}
\title{B-type anomaly coefficients for the D3-D5 domain wall}

\author{Marius de Leeuw,$^a$ Charlotte Kristjansen,$^b$ Georgios Linardopoulos,$^c$ and Matthias Volk$^d$}

\address[label1]{School of Mathematics \& Hamilton Mathematics Institute, Trinity College Dublin, Ireland \\
Trinity Quantum Alliance, Unit 16, Trinity Technology and Enterprise Centre, Pearse Street, Dublin 2, Ireland}
\address[label2]{Niels Bohr Institute, Copenhagen University, Blegdamsvej 17, 2100 Copenhagen \O, Denmark}
\address[label3]{Wigner Research Centre for Physics, Konkoly-Thege Mikl{ó}s \'{u}t 29-33, 1121 Budapest, Hungary}
\address[label4]{Institut f\"{u}r Physik, Humboldt-Universit\"{a}t zu Berlin, Zum Gro{\ss}en Windkanal 2, 12489 Berlin, Germany}

\begin{abstract}
We compute type-B Weyl anomaly coefficients for the domain wall version of $\N = 4$ SYM that is holographically dual to the D3-D5 probe-brane system with flux. Our starting point is the explicit expression for the improved energy momentum tensor of $\N = 4$ SYM. We determine the two-point function of this operator in the presence of the domain wall and extract the anomaly coefficients from the result. In the same process we determine the two-point function of the displacement operator.
\end{abstract}

\begin{keyword}
AdS/CFT correspondence, defect CFT, stress tensor, displacement operator, D3-D5 probe brane
\end{keyword}

\end{frontmatter}

\section{Introduction}
\noindent Tractable interacting boundary conformal field theories in four dimensions have been advertised for in the program which aims to study conformal anomalies in the presence of boundaries \cite{HerzogHuang17, HerzogHuangJensen17}. A class of such theories is constituted by the domain wall versions of $\N = 4$ SYM that are dual to the D3-D5 or the D3-D7 probe brane models with flux \cite{KarchRandall01a, KarchRandall01b, ConstableMyersTafjord99, MyersWapler08, KristjansenSemenoffYoung12b}. In these theories a subset of the scalar fields of $\N = 4$ SYM gets a non-zero vacuum expectation value (vev) in the form of a Nahm pole \cite{Nahm79c} on one side of a co-dimension one defect. In the D3-D5 case the classical solution conserves half of the supersymmetries of $\N = 4$ SYM \cite{GaiottoWitten08a} whereas in the D3-D7 case supersymmetry is lost \cite{MyersWapler08}. Both types of solutions preserve conformal symmetry on the defect itself, thus providing us with a defect conformal field theory (dCFT), and the presence of the non-vanishing vevs implies that numerous correlation functions are non-trivial already at tree level or at the level of a few contractions. \\
\indent These domain wall versions of $\N = 4$ SYM have been studied mainly for their integrability properties. The domain wall can be represented as a spin chain boundary state in the form of a matrix product state or a valence bond state and the computation of correlation functions (mainly one-point functions) can be formulated as the computation of the overlap between the boundary state and a Bethe eigenstate \cite{deLeeuwKristjansenZarembo15, Buhl-MortensenLeeuwKristjansenZarembo15, KristjansenMullerZarembo20a}. The D3-D5 domain wall has proven to be integrable at all loop orders \cite{Buhl-MortensenLeeuwIpsenKristjansenWilhelm17a, GomborBajnok20a, GomborBajnok20b, KomatsuWang20}. Conversely, one of the two D3-D7 domain wall setups seems to be integrable only at leading order \cite{deLeeuwGomborKristjansenLinardopoulosPozsgay19}, while the other is not integrable at all \cite{deLeeuwKristjansenVardinghus19}. Integrability has also been studied from the string theory perspective in \cite{DekelOz11b, LinardopoulosZarembo21}. For a review of these developments we refer to \cite{deLeeuwIpsenKristjansenWilhelm17, deLeeuw19, Linardopoulos20}. \\
\indent In the present letter, we advocate the D3-D5 domain wall model as a tractable interacting four dimensional boundary CFT where
anomaly coefficients of the stress tensor trace anomaly can be explicitly calculated and conjectured relations between them can be tested. \\
\indent An example of a relation between anomaly coefficients is the equality of the two type B anomaly coefficients for a supersymmetric co-dimension two defect CFT in four dimensions \cite{BianchiLemos19}. One of the coefficients appears in the one-point function of the stress tensor and the other in the two-point function of the displacement operator which hence are related. A similar relation between these two correlation functions is conjectured to hold in higher dimensions for defects of co-dimension larger than or equal to two \cite{BianchiLemos19}. \\
\indent For a defect of co-dimension one the one-point function of the stress tensor vanishes as only spin-less operators can have non-vanishing one-point functions in this case \cite{McAvityOsborn95, LiendoRastellivanRees12}. On the other hand, the two-point function of the displacement operator will in general be non-vanishing. For a flat co-dimension one defect the two-point function of the displacement operator can be extracted in a simple manner from the stress tensor two-point function. The stress tensor two-point function itself is very constrained by symmetries. More precisely, it is completely fixed up to three unknown functions of a certain conformal ratio \cite{McAvityOsborn93}. These functions encode information about the two-point function of the displacement operator as well as certain anomaly coefficients of the theory. One of the unknown functions in a certain limit becomes equal to the B-type anomaly coefficient of the $\K \W$ term, denoted as $b_2$,\footnote{Here $\K$ refers to the extrinsic curvature of the defect and $\W$ to the Weyl tensor of the background.} and in another limit to the anomaly coefficient denoted as $c$ \cite{HerzogHuang17}. It was noticed that for free theories $b = 8c$ \cite{Fursaev15}. In our domain wall model this relation is not fulfilled. \\
\indent We determine explicitly the three unknown functions entering the two-point function of the stress tensor for the D3-D5 defect CFT at the leading order in the 't Hooft coupling. This allows us to determine the above mentioned anomaly coefficient and the two-point function of the displacement operator. \\
\indent Our letter is organized as follows. We start by introducing in more precise terms the domain wall version of $\N = 4$ dual to the D3-D5 probe brane system with flux in section \ref{Section:DomainWall}. In section \ref{StressTensor} we express the improved stress tensor of $\N = 4$ SYM explicitly in terms of the fundamental fields of the theory. Subsequently, in sections \ref{Section:TwoPoint} and \ref{Section:Displacement} we derive respectively the two-point function of the stress tensor in the presence of the domain wall and the two-point function of the associated displacement operator. Finally, section \ref{Section:Conclusion} contains our conclusions. Certain definitions as well as various details of our computations can be found in \ref{Appendix:ColorDecomposition}--\ref{Appendix:PropagatorsD3D5}.
\section{The D3-D5 domain wall\label{Section:DomainWall}}
\noindent We start from the Lagrangian density of $\N=4$ SYM theory which reads:\footnote{We follow the conventions of \cite{Buhl-MortensenLeeuwIpsenKristjansenWilhelm16c} by adopting the mostly-plus signature $(-+++)$ for the Minkowski metric.}
\begin{IEEEeqnarray}{ll}
\LL_{\N = 4} = &\frac{2}{g_{\text{\scalebox{.8}{YM}}}^2} \text{tr}\bigg\{-\frac{1}{4} F_{\mu\nu} F^{\mu\nu} - \frac{1}{2} \left(D_{\mu}\varphi_i\right)^2 + i\,\bar{\psi}_{\alpha}\slashed{D}\,\psi_{\alpha} + \frac{1}{4}\left[\varphi_i,\varphi_j\right]^2 \nonumber \\
&+ \sum_{i = 1}^{3}G^i_{\alpha\beta}\bar{\psi}_{\alpha}\left[\varphi_i,\psi_{\beta}\right] + \sum_{i = 4}^{6}G^i_{\alpha\beta} \bar{\psi}_{\alpha} \gamma_5 \left[\varphi_i,\psi_{\beta}\right]\bigg\}, \qquad \label{LagrangianSYM}
\end{IEEEeqnarray}
where $\bar{\psi}_{\alpha} \equiv \psi_{\alpha}^{\dagger} \gamma^0$, $\slashed{D} \equiv \gamma^{\mu}D_{\mu}$ and
\begin{IEEEeqnarray}{l}
F_{\mu\nu} \equiv \partial_{\mu}A_{\nu} - \partial_{\nu}A_{\mu} - i \left[A_{\mu},A_{\nu}\right], \quad D_{\mu}f \equiv \partial_{\mu}f - i \left[A_{\mu},f\right], \qquad \label{CovariantDerivatives}
\end{IEEEeqnarray}
and the definitions of the matrices $\gamma^{\mu}$ and $G^{i}$ can be found in \cite{Buhl-MortensenLeeuwIpsenKristjansenWilhelm16c}. All the fields (gluons, scalars, fermions) of the Lagrangian \eqref{LagrangianSYM} carry adjoint $\mathfrak{u}(N)$ color indices as follows:
\begin{IEEEeqnarray}{l}
A_{\mu} = A_{\mu}^{a}T^a, \qquad \varphi_{i} = \varphi_{i}^{a}T^a, \qquad \psi_{\alpha} = \psi_{\alpha}^{a}T^a, \qquad \label{ColorDecompositionSYM}
\end{IEEEeqnarray}
where $a = 1,\ldots, N^2$, $\mu = 0,\ldots,3$, $i = 1,\ldots,6$ and $\alpha = 1,\ldots,4$. The basic properties of the $N \times N$ generators $T^a$ of $\mathfrak{u}(N)$ have been collected in \ref{Appendix:ColorDecomposition}. The details of the convention for the spinors, which are of Majorana-Weyl type, can be found in \cite{Buhl-MortensenLeeuwIpsenKristjansenWilhelm16c}, appendix C. The equations of motion that follow from the action \eqref{LagrangianSYM} are:
\begin{IEEEeqnarray}{c}
D^{\mu}F_{\mu\nu} = i\left[D_{\nu}\varphi_i,\varphi_i\right], \qquad D^{\mu}D_{\mu}\varphi_i = \left[\varphi_j,\left[\varphi_j,\varphi_i\right]\right], \\
i\slashed{D}\psi_{\alpha} = \sum_{i=1}^{3}G_{\alpha\beta}^{i}\left[\psi_{\beta},\varphi_i\right] + \sum_{i=4}^{6} G_{\alpha\beta}^{i}\gamma_5\left[\psi_{\beta},\varphi_i\right].
\end{IEEEeqnarray}
These equations have a number of ``fuzzy funnel'' solutions which break the PSU(2,2$|$4) global symmetry of $\N = 4$ SYM keeping either 1/2 supersymmetry (D3-D5 case) or none (D3-D7 case). The simplest and most studied of these constitute the dual of the D3-D5 probe brane system with flux and reads with $i=1,2,3$
\begin{IEEEeqnarray}{l}
A_{\mu} = 0, \hspace{0.5cm}\psi_{\alpha} = 0,\\
\varphi_{i} = \varphi_{i}^{\text{cl}}\left(x_3\right) = \frac{1}{x_3} \cdot \left[\begin{array}{cc} \left(t_i\right)_{k\times k} & 0_{k\times \left(N - k\right)} \\ 0_{\left(N - k\right)\times k} & 0_{\left(N - k\right)\times \left(N - k\right)} \end{array}\right]_{N\times N} \\
\varphi_{i+3} = 0, \qquad \label{FuzzyFunnelD3D5}
\end{IEEEeqnarray}
where the matrices $t_i$ furnish a $k$-dimensional irreducible representation of $\mathfrak{su}\left(2\right)$:
\begin{IEEEeqnarray}{c}
\left[t_i, t_j\right] = i \epsilon_{ijl}t_l, \qquad i,j,l = 1,2,3,
\end{IEEEeqnarray}
and where we will assume that $x_3 > 0$. Furthermore, we will assume that $k \geq 2$. We shall briefly comment on the special cases $k = 1$ and $k = 0$ later. In the string theory dual $k$ is a flux parameter. Quantizing around this classical solution one finds that only the field components in the $(N-k) \times (N-k)$ block of the fields that are present in the $x_3 > 0$ region can propagate to the $x_3 < 0$ region. This way, we effectively get a domain wall set-up where the gauge group of the theory is $U(N-k)$ for $x_3 < 0$ and (broken) $U(N)$ for $x_3 > 0$, with the full $U(N)$ gauge symmetry only being recovered as $x_3\rightarrow \infty$. The set-up obviously breaks translational invariance in the $x_3$ direction but conformal symmetry is conserved on the $x_3 = 0$ domain wall leaving us with a Osp(4$|$4) symmetric dCFT.
\section{The improved stress tensor of ${\cal N}=4$ SYM\label{StressTensor}}
\noindent The stress tensor of $\N = 4$ SYM can be computed from the action \eqref{LagrangianSYM} by the canonical prescription,\footnote{Alternatively, we may use the covariant prescription (see e.g.\ \cite{BirrellDavies99})
\begin{IEEEeqnarray}{ll}
T_{\mu\nu} \equiv \frac{2}{\sqrt{-g}}\cdot\frac{\delta S}{\delta g^{\mu\nu}}, \qquad S = \int dx^4 \sqrt{-g} \, \LL, \label{StressTensorCovariant}
\end{IEEEeqnarray}
which leads to a manifestly symmetric stress tensor. Note however that the fermionic terms of the action must be varied with respect to the vierbein field $e_{\mu}^m$ instead of the graviton $g^{\mu\nu} = e^{\mu}_m e^{\nu m}$.}
\begin{IEEEeqnarray}{ll}
T_{\mu\nu} = &\frac{\partial\LL}{\partial\partial^{\mu}A_{\rho}}\,\partial_{\nu}A_{\rho} + \frac{\partial\LL}{\partial\partial^{\mu}\varphi_{i}}\,\partial_{\nu}\varphi_{i} + \frac{\partial\LL}{\partial\partial^{\mu}\bar{\psi}_{\alpha}}\,\partial_{\nu}\bar{\psi}_{\alpha} + \nonumber \\
& + \frac{\partial\LL}{\partial\partial^{\mu}\psi_{\alpha}}\,\partial_{\nu}\psi_{\alpha} - g_{\mu\nu}\LL, \label{StressTensorCanonical}
\end{IEEEeqnarray}
but it is neither symmetric, nor traceless, nor conserved. Following Callan, Coleman and Jackiw \cite{CallanColemanJackiw70}, an improved version of the stress tensor can be constructed by applying a series of transformations to the canonical formula \eqref{StressTensorCanonical},
\begin{IEEEeqnarray}{ll}
\Theta_{\mu\nu} = \frac{2}{g_{\text{\scalebox{.8}{YM}}}^2} \cdot \text{tr}\bigg\{&-{F_{\mu}}^{\varrho} F_{\nu\varrho} - \frac{2}{3} \left(D_{\mu}\varphi_i\right) \Bigl(D_{\nu}\varphi_i\Bigr) + \frac{1}{3} \, \varphi_i \, D_{(\mu} D_{\nu)}\varphi_i + \nonumber \\
& + \frac{i}{2} \, \bar{\psi}_{\alpha}\gamma_{(\mu}\overset{\leftrightarrow}{D}_{\nu)}\psi_{\alpha}\bigg\} - g_{\mu\nu} \Lambda, \qquad \label{StressTensorImprovedSYM}
\end{IEEEeqnarray}
where $a_{(\mu\nu)} \equiv \left(a_{\mu\nu} + a_{\nu\mu}\right)/2$ and $f\overset{\leftrightarrow}{\partial}_{\mu} g \equiv f \left(\partial_{\mu}g\right) - \left(\partial_{\mu}f\right) g$. Furthermore,
\begin{IEEEeqnarray}{ll}
\Lambda \equiv \frac{2}{g_{\text{\scalebox{.8}{YM}}}^2} \cdot \text{tr}\left\{-\frac{1}{4} \, F_{\mu\nu} F^{\mu\nu} - \frac{1}{6} \left(D_{\mu}\varphi_i\right)^2 - \frac{1}{12}\left[\varphi_i,\varphi_j\right]^2\right\}. \qquad
\end{IEEEeqnarray}
Details are provided in \ref{Appendix:ImprovedStressTensor}. Apart from being manifestly symmetric, the stress tensor \eqref{StressTensorImprovedSYM} is also on-shell traceless and conserved:
\begin{IEEEeqnarray}{ll}
\Theta_{\mu\nu} = \Theta_{\nu\mu}, \qquad g^{\mu\nu} \Theta_{\mu\nu} = 0, \qquad D^{\mu}\Theta_{\mu\nu} = 0. \qquad
\end{IEEEeqnarray}
The (unregularized) 2-point function of the improved stress tensor can be written in terms of the inversion tensors $I_{\mu\nu}$, $\I_{\mu\nu\rho\sigma}$ \cite{OsbornPetkou93}:
\begin{IEEEeqnarray}{c}
\left<\Theta_{\mu\nu}\left(x\right) \Theta_{\rho\sigma}\left(y\right)\right>_{\text{bare}} = \frac{C_{T}}{4 \pi^4 s^8} \cdot \I_{\mu\nu\rho\sigma} \label{BareStressTensorSYM1}, \\[6pt]
\I_{\mu\nu\rho\sigma} \equiv \frac{1}{2}\Big[I_{\mu\rho}I_{\nu\sigma} + I_{\mu\sigma}I_{\nu\rho} - \frac{1}{2}\,g_{\mu\nu}g_{\rho\sigma}\Big], \ I_{\mu\nu} \equiv g_{\mu\nu} - 2\frac{{s}_{\mu} {s}_{\nu}}{s^2}, \qquad \label{BareStressTensorSYM2}\\[6pt]
s_{\mu} \equiv x_{\mu} - y_{\mu}, \quad s^2 \equiv s_{\mu}s^{\mu}.
\end{IEEEeqnarray}
The coefficient $C_{T}$ is equal to
\begin{IEEEeqnarray}{c}
C_{T} = 16n_{\text{g}} + \frac{4}{3}n_{\text{s}} + 4n_{\text{f}} = 40N^2, \label{BareStressTensorSYMcoefficient}
\end{IEEEeqnarray}
where $n_{\text{g}}$, $n_{\text{s}}$, $n_{\text{f}}$ provide the numbers of gluon, scalar and Majorana-Weyl fields respectively \cite{Osborn19} and where $N$ is the rank of the gauge group.\footnote{Notice that for our interface set up the rank of the gauge group for $x_3 <0$ is $(N-k)$.} A regularized version of \eqref{BareStressTensorSYM1}--\eqref{BareStressTensorSYM2} is also available \cite{ErdmengerOsborn96}. The 2 and 3-point functions of the stress tensor are protected within $\N = 4$ SYM (see e.g.\ \cite{HoweSokatchevWest98} and references therein).
\section{Stress tensor two-point function of the defect CFT\label{Section:TwoPoint}}
\noindent The stress tensor one-point function of the defect CFT can be directly computed by plugging the fuzzy-funnel solution \eqref{FuzzyFunnelD3D5} into the formula \eqref{StressTensorImprovedSYM}. As only scalars have non-zero vevs it is enough to consider the contribution to the stress tensor coming from scalars which reads
\begin{IEEEeqnarray}{ll}
\Theta_{\mu\nu \text{(scalars)}} = \frac{2}{g_{\text{\scalebox{.8}{YM}}}^2} \cdot \text{tr}\bigg\{&-\frac{2}{3}\left(\partial_{\mu}\varphi_i\right)\Bigl(\partial_{\nu}\varphi_i\Bigr) + \frac{1}{3}\,\varphi_i\left(\partial_{\mu}\partial_{\nu}\varphi_i\right) + \nonumber \\
& + \frac{1}{6} \, g_{\mu\nu}\left[\left(\partial_{\varrho}\varphi_i\right)^2 + \frac{1}{2}\left[\varphi_i,\varphi_j\right]^2\right]\bigg\}, \qquad \label{StressTensorScalar}
\end{IEEEeqnarray}
and which is also obviously symmetric, on-shell traceless and conserved. We find that the classical value of the stress tensor vanishes
\begin{IEEEeqnarray}{ll}
\Theta_{\mu\nu}^{\text{cl}} = 0, \label{StressTensorOnePointFunction}
\end{IEEEeqnarray}
as it has to for a co-dimension one defect \cite{McAvityOsborn93, McAvityOsborn95}. The leading contribution to the connected part of the stress tensor two-point function consists of a single Wick contraction and is of order $g_{\text{\scalebox{.8}{YM}}}^{-2}$. More specifically
\begin{IEEEeqnarray}{c}
\left<\Theta_{\mu\nu}\left(x\right)\Theta_{\rho\sigma}\left(y\right)\right> =
N\left(
\scriptsize{\feynmandiagram [small, inline =(a.base), horizontal= a to b]
{a [dot] -- [blue, edge label=\(\color{black}\lambda^{-1}\)] b [dot]};} + \scriptsize{\feynmandiagram [small, inline =(a.base), horizontal= a to b]
{a [dot] -- [blue, out=45, in=135, edge label=\(\color{black}\lambda^0\)] b [dot] -- [blue, out=225, in=315] a};} + \ldots \right),
\label{StressTensorTwoPointFunctionD3D5}
\end{IEEEeqnarray}
where the 't Hooft coupling is defined as $\lambda=g_{\text{\scalebox{.8}{YM}}}^{2}N$. Expanding the fields around the classical solution \eqref{FuzzyFunnelD3D5},
\begin{IEEEeqnarray}{c}
A_{\mu} = \tilde{A}_{\mu}, \qquad \psi_{\alpha} = \tilde{\psi}_{\alpha}, \qquad \varphi_i\left(x\right) = \varphi_i^{\text{cl}}\left(x_3\right) + \tilde{\varphi}_i\left(x\right), \qquad \label{FieldPerturbationD3D5}
\end{IEEEeqnarray}
for $i = 1,\ldots,6$, it follows that the leading correction to the stress tensor \eqref{StressTensorImprovedSYM} takes the form:
\begin{IEEEeqnarray}{ll}
\Theta_{\mu\nu}^{(1)}\left(x\right) = \frac{1}{g_{\text{\scalebox{.8}{YM}}}^2} \, \frac{4}{3 x_3^2} \, \text{tr}\bigg\{&\Big(\frac{1}{x_3} \left(n_{\mu} n_{\nu} - g_{\mu\nu}\right) \tilde{\varphi}_i + n_{\mu} \partial_{\nu}\tilde{\varphi}_i + n_{\nu} \partial_{\mu}\tilde{\varphi}_i - \nonumber \\
& - \frac{g_{\mu\nu}}{2} \, \partial_{3}\tilde{\varphi}_i + \frac{x_3}{2} \, \partial_{\mu}\partial_{\nu}\tilde{\varphi}_i\Big) t_i\bigg\}, \qquad \label{StressTensorFluctuationsD3D5}
\end{IEEEeqnarray}
where $n_{\mu} \equiv \delta_{\mu3}$ denotes the normal vector to the defect. The expression \eqref{StressTensorFluctuationsD3D5} can be inserted into \eqref{StressTensorTwoPointFunctionD3D5}. The two-point function only obtains contributions from the $k\times k$ block of the quantum fields. This block is conveniently expanded in terms of the fuzzy $\mathfrak{su}\left(2\right)$ spherical harmonics as explained in \cite{Buhl-MortensenLeeuwIpsenKristjansenWilhelm16a, Buhl-MortensenLeeuwIpsenKristjansenWilhelm16c} and summarized in \ref{Appendix:PropagatorsD3D5}. The expression for the bulk two-point function of the stress tensor \eqref{StressTensorTwoPointFunctionD3D5} contains the Wick-contracted quantity
\begin{IEEEeqnarray}{ll}
\wick{\text{tr}[t_i \c1{\tilde{\varphi}_i}] \cdot \text{tr}[t_j \c1{\tilde{\varphi}_j}]} &= c_k \cdot K^{5/2}\left(x,y\right) = \nonumber \\
& = c_k \cdot \frac{g_{\text{\scalebox{.8}{YM}}}^2}{320\pi^2} \frac{1}{x_3 y_3} \frac{{}_2 F_1\left(2,3,6 ; -\xi^{-1}\right)}{\xi^3\left(1 + \xi\right)} \qquad \label{ScalarPropagatorD3D5}
\end{IEEEeqnarray}
and its derivatives. The conformal ratios $\xi, v$ and the constants $c_k$ are defined as:
\begin{IEEEeqnarray}{ll}
\xi \equiv \frac{s^2}{4x_3y_3}, \qquad v^2 \equiv \frac{\xi}{1 + \xi}, \qquad c_k \equiv \frac{k\left(k^2 - 1\right)}{4}.
\end{IEEEeqnarray}
The expression \eqref{ScalarPropagatorD3D5} is obtained by using the scalar propagator of the references \cite{Buhl-MortensenLeeuwIpsenKristjansenWilhelm16c, deLeeuwIpsenKristjansenVardinghusWilhelm17} (see \eqref{ScalarPropagatorD3D5complicated} in \ref{Appendix:PropagatorsD3D5}) and the color factors \eqref{ColorFactor1}--\eqref{ColorFactor2}. Performing all contractions we arrive at the following result
\begin{IEEEeqnarray}{ll}
\scriptsize{\feynmandiagram [small, inline =(a.base), horizontal= a to b]
{a [dot] -- [blue, edge label=\(\color{black}\lambda^{-1}\)] b [dot]};} = \left<\Theta_{\mu\nu}^{(1)}\left(x\right)\Theta_{\rho\sigma}^{(1)}\left(y\right)\right> = \\
= \frac{1}{s^8} \cdot \Bigg\{\left(X_{\mu}X_{\nu} - \frac{g_{\mu\nu}}{4}\right)\left(Y_{\rho}Y_{\sigma} - \frac{g_{\rho\sigma}}{4}\right) A\left(v\right) + \Big(X_{\mu}Y_{\rho}I_{\nu\sigma} + \nonumber \\
+ X_{\mu}Y_{\sigma}I_{\nu\rho} + X_{\nu}Y_{\sigma}I_{\mu\rho} + X_{\nu}Y_{\rho}I_{\mu\sigma} - g_{\mu\nu}Y_{\rho}Y_{\sigma} - g_{\rho\sigma}X_{\mu}X_{\nu} + \nonumber \\
+ \frac{1}{4}\,g_{\mu\nu}g_{\rho\sigma}\Big) B\left(v\right) + \I_{\mu\nu\rho\sigma} C\left(v\right)\Bigg\}, \qquad \label{StressTensorTwoPointFunction}
\end{IEEEeqnarray}
which has the generic form specified in \cite{McAvityOsborn93, McAvityOsborn95} where
\begin{IEEEeqnarray}{l}
X_{\mu} \equiv v\left(\frac{2x_3}{s^2}\,s_{\mu} - n_{\mu}\right), \quad Y_{\rho} \equiv -v\left(\frac{2y_3}{s^2}\,s_{\rho} + n_{\rho}\right). \qquad
\end{IEEEeqnarray}
Symmetry arguments fix the two-point function up to the three functions $A(\nu)$, $B(\nu)$ and $C(\nu)$. The functions are not independent but fulfill the relations \cite{McAvityOsborn95}
\begin{IEEEeqnarray}{l}
\left(v\frac{d}{dv} - 4\right)\left(C + 2B\right) = -\frac{1}{2}\left(A + 4B\right) - 4C, \\
\left(v\frac{d}{dv} - 4\right)\left(3A + 4B\right) = 2A - 24B.
\end{IEEEeqnarray}
For our domain wall set-up these functions take the form
\begin{IEEEeqnarray}{l}
A\left(v\right) = 4 \gamma \left(6v^6 + 3v^4 + v^2\right) \label{ConstantA}, \\
B\left(v\right) = - \gamma\left(3v^6 - v^4 - 2v^2\right) \label{ConstantB}, \\
C\left(v\right) = \gamma v^2 \left(v^2 - 1\right)^2, \qquad \label{ConstantC}
\end{IEEEeqnarray}
where
\begin{IEEEeqnarray}{c}
\gamma \equiv \frac{32 c_k}{9 \pi^2 g_{\text{\scalebox{.8}{YM}}}^2}. \quad \label{alpha}
\end{IEEEeqnarray}
Far away from the boundary we recover the $\N = 4$ two-point function which vanishes at the linearized level and is only nonzero at quadratic order \eqref{BareStressTensorSYM2}. Indeed, the functions $A$, $B$, $C$ and the stress tensor two-point function \eqref{StressTensorTwoPointFunction} all vanish for $\xi$, $v \rightarrow 0$. \\
\indent Our derivation of the two-point function of the stress tensor is strictly speaking only valid for $k \geq 2$ but the result indicates that the correlation function should vanish for the special cases $k = 0,1$. This can indeed be seen to be the case. For $k = 1$ there are no vevs, and instead one has to by hand impose boundary conditions on the $1 \times (N-1)$, $(N-1) \times 1$ and $1\times 1$ blocks of all fields, Dirichlet or Neumann, following a certain recipe that ensures supersymmetry \cite{deLeeuwIpsenKristjansenVardinghusWilhelm17, KristjansenMullerZarembo20a}. In this case the leading contribution to the two-point function involves two contractions and is of higher order in $\lambda$ than the one for $k \geq 2$. The case $k = 0$ is a completely different set-up where there are no particular boundary conditions on the bulk fields but additional fundamental fields living on the defect and interacting among themselves and with the bulk fields \cite{DeWolfeFreedmanOoguri01}. Also in this case the two-point function of the stress tensor involves two contractions and is of higher order in $\lambda$. \\
\indent In \cite{HerzogHuang17} the anomaly coefficient of the $\K\W$ term, denoted as $b_2$, was argued to be given by the expression
\begin{equation}
b_2 = \frac{2\pi^4}{15} \cdot \alpha(1),
\end{equation}
with
\begin{equation}
\alpha(\nu) = \frac{d-1}{d^2} \cdot \left[(d-1)(A(\nu) + 4B(\nu))+ d C(\nu)\right],
\end{equation}
and with $d$ being the dimension, i.e.\ $d = 4$ in our case. Making use of the relations \eqref{ConstantA}--\eqref{ConstantC} we find that for the D3-D5 domain wall
\begin{equation}
\alpha(1) = \frac{9}{16} \cdot A(1) = \frac{20 k\,(k^2-1)}{\pi^2 g_{\text{\scalebox{.8}{YM}}}^2},
\quad b_2= \frac{8 \pi^2}{3 g_{\text{\scalebox{.8}{YM}}}^2} \cdot k\,(k^2-1). \label{alpha1}
\end{equation}
It has been observed that for free theories $\alpha(1)=2\alpha(0)$ or equivalently $b_2=8c$ \cite{Fursaev15, HerzogHuang17}. This relation is clearly not fulfilled in our interacting case for $k\geq 2$ where (at leading order) $\alpha(0)=0$ and $\alpha(1)\neq 0$.
\section{Displacement operator\label{Section:Displacement}}
\noindent The displacement operator $\D$ can be defined from the divergence of the improved (scalar) stress tensor \eqref{StressTensorScalar} as follows:
\begin{IEEEeqnarray}{ll}
\partial^{\mu}\Theta_{\mu\nu} = \delta(x_3) \,\eta_{\nu}\,\D, \label{StressTensorDivergence}
\end{IEEEeqnarray}
where $\eta_{\mu} \equiv \left(0,0,0,1\right)$ is the unit normal to the $x_3= 0$ plane boundary. Integrating over $x_3$ from $0^-$ to $0^+$ and bearing in mind the conformal invariance of the defect we find
\begin{IEEEeqnarray}{ll}
\D\left(\textbf{x}\right) = \lim_{x_3\rightarrow0+}\Theta_{33}\left(\textbf{x},x_3\right)-\lim_{x_3\rightarrow0-}\Theta_{33}\left(\textbf{x},x_3\right),
\quad
\end{IEEEeqnarray}
with $\textbf{x}\equiv \left(x_0,x_1,x_2\right)$. The leading order contribution to the two-point function of the displacement operator can thus can be directly read off from \eqref{StressTensorTwoPointFunction}:
\begin{IEEEeqnarray}{c}
\left<\D^{(1)}\left(\textbf{x}\right)\D^{(1)}\left(\textbf{y}\right)\right> = \lim_{x_3,y_3\rightarrow0^+}\left<\Theta^{(1)}_{33}\left(x\right)\Theta^{(1)}_{33}\left(y\right)\right>
=\frac{\alpha(1)}{s^8}, \qquad
\end{IEEEeqnarray}
where $\alpha(1)$ was given in (\ref{alpha1}).
The terms involving $\lim x_3\rightarrow 0^-$ are of higher order in $\lambda$.
\section{Conclusion \label{Section:Conclusion}}
\noindent We have illustrated that the D3-D5 domain wall version of ${\cal N}=4$ SYM is a tractable, interacting four dimensional co-dimension one defect CFT where correlation functions involving the stress tensor and hence certain anomaly coefficients can be explicitly calculated. So far we evaluated the two-point function of the stress tensor as well as its associated displacement operator and extracted the coefficient $b_2$. In particular, we found that $b_2\neq 8c$ for this theory. It might be interesting to compute the three-point function of the stress tensor and extract the other B-type anomaly coefficient for this case, $b_1$ \cite{HerzogHuang17}. \\
\indent The present defect CFT should also be amenable to the boundary conformal bootstrap program. Thus one could imagine that the explicit perturbative calculation of the leading order contribution to the two-point functions as pursued in this paper in combination with the all loop results for the one-point functions \cite{GomborBajnok20a, GomborBajnok20b} could be used as input to start a bootstrap procedure to determine higher loop correlation functions. First steps in the direction of developing such a program were taken in \cite{deLeeuwIpsenKristjansenVardinghusWilhelm17}. \\
\indent Correlation functions of the D3-D5 and D3-D7 domain wall versions of ${\cal N}=4$ SYM can be studied from the strong coupling string perspective as well. One-point functions of chiral primaries can be explicitly matched between gauge theory and string theory using supersymmetric localization \cite{KomatsuWang20} and show, for an operator of length $2l$, a scaling $k^{2l+1}/\lambda^l$ for weak coupling and large flux \cite{NagasakiYamaguchi12}. It would be interesting if one could reproduce from holography the weak coupling, large flux scaling of the two-point function observed here and in \cite{deLeeuwIpsenKristjansenVardinghusWilhelm17}. The holographic computation of two-point functions for the specific D3-D5 probe brane set-up considered here was addressed in \cite{GeorgiouLinardopoulosZoakos23}. A computation of holographic two-point functions for generic Karch-Randall set-up's, not specifying the brane or supergravity action, has been pursued in e.g.\ \cite{RastelliZhou17a, GoncalvesItsios18, KavirajPaulos18, MazacRastelliZhou18, KastikainenShashi21, ChenZhou23, GimenezGrau23}.
\section*{Acknowledgments}
\noindent We thank Jonah Baerman, Lorenzo Bianchi, Adam Chalabi, Pedro Liendo, Ioannis Papadimitriou, Anton Pribytok, Volker Schomerus, and Konstantin Zarembo for discussions. M.\ de L.\ and C.K.\ thank KITP, Santa Barbara for hospitality and for support in part of this research by the National Science Foundation under Grant No. NSF PHY-1748958. M.\ de L.\ was supported by SFI and the Royal Society for funding under grants UF160578, RGF$\backslash$R1$\backslash$181011, RGF$\backslash$EA$\backslash$180167 and RF$\backslash$ERE$\backslash$210373. C.K.\ was in addition supported in part by DFF-FNU through grant number 1026-00103B. The work of G.L.\ was supported by the National Development Research and Innovation Office (NKFIH) research grant K134946.
\appendix\section{Color decompositions \label{Appendix:ColorDecomposition}}
\noindent All the fields of $\N = 4$ SYM carry adjoint $\mathfrak{u}(N)$ color indices \eqref{ColorDecompositionSYM}, where $\mathfrak{u}(N) = \mathfrak{su}(N) \times \mathfrak{u}(1)$. The fundamental representation of $\mathfrak{su}(N)$ is spanned by of $N^2-1$ traceless hermitian $N \times N$ generators. These are normalized as
\begin{IEEEeqnarray}{l}
\text{tr}\left[T^a T^b\right] = \delta^{ab}, \label{ColorNormalization}
\end{IEEEeqnarray}
while they also satisfy the $\mathfrak{su}(N)$ Fierz identity:
\begin{IEEEeqnarray}{l}
\left(T^a\right)_{\mathfrak{m}\mathfrak{n}} \left(T^a\right)_{\mathfrak{r}\mathfrak{s}} = \delta_{\mathfrak{m}\mathfrak{s}} \delta_{\mathfrak{n}\mathfrak{r}} - \frac{1}{N}\,\delta_{\mathfrak{m}\mathfrak{n}} \delta_{\mathfrak{r}\mathfrak{s}}. \label{FierzIdentitySU}
\end{IEEEeqnarray}
The extra $\mathfrak{u}(1)$ generator that is needed to obtain the fundamental representation of $\mathfrak{u}(N)$ is proportional to the identity matrix:
\begin{IEEEeqnarray}{l}
\left(T^{0}\right)_{\mathfrak{m}\mathfrak{n}} = \frac{1}{\sqrt{N}}\,\delta_{\mathfrak{m}\mathfrak{n}}.
\end{IEEEeqnarray}
The $N^2$ hermitian generators of $\mathfrak{u}(N)$ still satisfy \eqref{ColorNormalization}, whereas the $\mathfrak{u}(N)$ Fierz identity becomes:
\begin{IEEEeqnarray}{l}
\left(T^a\right)_{\mathfrak{m}\mathfrak{n}} \left(T^a\right)_{\mathfrak{r}\mathfrak{s}} = \delta_{\mathfrak{m}\mathfrak{s}} \delta_{\mathfrak{n}\mathfrak{r}}. \label{FierzIdentityU}
\end{IEEEeqnarray}
\section{Improved stress tensor \label{Appendix:ImprovedStressTensor}}
\noindent The recipe for computing the improved stress tensor can be found in e.g.\ \cite{Osborn19}. It can be adapted to a covariant context as follows:
\begin{IEEEeqnarray}{ll}
\Theta_{\mu\nu} = & T_{\mu\nu} + D^{\rho}\left(X_{\rho\mu\nu} - X_{\mu\rho\nu} - X_{\nu\rho\mu} + Y_{\rho\mu\nu} - Y_{\mu\nu\rho} - Y_{\nu\mu\rho}\right) + \nonumber \\[6pt]
& + \mathcal{D}_{\mu\nu\rho\sigma}Z^{\rho\sigma}, \label{StressTensorImproved}
\end{IEEEeqnarray}
where $T_{\mu\nu}$ is the stress tensor \eqref{StressTensorCovariant} or \eqref{StressTensorCanonical}, and $X$ and $Z$ must be determined so that
\begin{IEEEeqnarray}{l}
T_{[\mu\nu]} = -D^{\rho} X_{\rho\mu\nu}, \qquad X_{\rho\mu\nu} = -X_{\rho\nu\mu} \\[6pt]
D^{\mu}T_{\mu\nu} = D^{\rho}D^{\mu}Y_{\nu\rho\mu}, \qquad Y_{\rho\mu\nu} = Y_{\rho\nu\mu} \\[6pt]
T^{\mu}_{\mu} = 2D^{\mu} X_{\rho\mu\nu} g^{\rho\nu} + 2D^{\mu} Y_{\rho\nu\mu} g^{\rho\nu} - D^{\mu} Y_{\mu\rho\nu} g^{\rho\nu} + D^{\mu}D^{\nu}Z_{\mu\nu} \qquad \\[6pt]
Z_{\mu\nu} = Z_{\nu\mu}
\end{IEEEeqnarray}
and $\mathcal{D}_{\mu\nu\rho\sigma}$ is defined in $d$-dimensional spacetime as
\begin{IEEEeqnarray}{l}
\mathcal{D}_{\mu\nu\rho\sigma} = \frac{1}{d-2}\Big(g_{\mu(\rho} D_{\sigma)}D_{\nu} + g_{\nu(\rho} D_{\sigma)} D_{\mu} - g_{\mu(\rho}g_{\sigma)\nu} D^2 - \nonumber \\
\hspace{1cm} - g_{\mu\nu} D_{\rho}D_{\sigma}\Big) - \frac{1}{(d-1)(d-2)}\left(D_{\mu}D_{\nu} - g_{\mu\nu}D^2\right)g_{\rho\sigma}. \qquad
\end{IEEEeqnarray}
We have also used the definitions
\begin{IEEEeqnarray}{ll}
a_{(\mu\nu)} \equiv \frac{a_{\mu\nu} + a_{\nu\mu}}{2}, \qquad a_{[\mu\nu]} \equiv \frac{a_{\mu\nu} - a_{\nu\mu}}{2}.
\end{IEEEeqnarray}
By construction, the improved stress tensor \eqref{StressTensorImproved} is on-shell conserved, symmetric and traceless:
\begin{IEEEeqnarray}{ll}
D^{\mu}\Theta_{\mu\nu} = 0, \qquad \Theta_{[\mu\nu]} = 0, \qquad g^{\mu\nu}\Theta_{\mu\nu} = 0.
\end{IEEEeqnarray}
\section{Scalar propagators in the D3-D5 dCFT \label{Appendix:PropagatorsD3D5}}
\noindent The fluctuations of the upper $k \times k$ blocks of the scalar fields are expanded in fuzzy $\mathfrak{su}\left(2\right)$ spherical harmonics as follows:
\begin{IEEEeqnarray}{ll}
\left[\tilde{\varphi}_i\right]_{n_1,n_2} = \sum_{\ell = 1}^{k-1}\sum_{m = -\ell}^{\ell}\left[\hat{Y}^{m}_{\ell}\right]_{n_1,n_2} (\tilde{\varphi}_i)_{\ell,m}, \quad n_{1,2} = 1,\ldots,k. \qquad \label{FuzzySphericalHarmonicDecomposition}
\end{IEEEeqnarray}
The corresponding propagators are given by \cite{Buhl-MortensenLeeuwIpsenKristjansenWilhelm16c}:
\begin{IEEEeqnarray}{l}
\big<(\tilde{\varphi}_i)_{\ell m} (\tilde{\varphi}_j)_{\ell' m'}\big> = \left(-1\right)^{m'} \delta_{ij}\delta_{\ell\ell'} \delta_{m+m'} \times \bigg(\frac{\ell + 1}{2\ell + 1}K^{\nu = \ell - \frac{1}{2}} + \nonumber \\[6pt]
+ \frac{\ell}{2\ell + 1}K^{\nu = \ell + \frac{3}{2}}\bigg) - i\left(-1\right)^{m'}\delta_{\ell \ell'}\epsilon_{ijl}\left[t^{\left(2\ell + 1\right)}_{l}\right]_{\ell-m+1,\ell+m'+1} \times \nonumber \\[6pt]
\times \frac{1}{2\ell + 1} \left(K^{\nu = \ell - \frac{1}{2}} - K^{\nu = \ell + \frac{3}{2}}\right) \qquad \label{ScalarPropagatorD3D5complicated} \\[12pt]
\big<(\tilde{\varphi}_{i+3})_{\ell m} (\tilde{\varphi}_{j+3})_{\ell' m'}\big> = \left(-1\right)^{m'} \delta_{ij}\delta_{\ell\ell'} \delta_{m+m'} \times K^{\nu = \ell + \frac{1}{2}}, \label{ScalarPropagatorD3D5easy}
\end{IEEEeqnarray}
where $i,j,l = 1,2,3$ and \cite{deLeeuwIpsenKristjansenVardinghusWilhelm17}
\begin{IEEEeqnarray}{l}
K^{\nu}\left(x,y\right) = \frac{g_{\text{\scalebox{.8}{YM}}}^2}{16\pi^2}\frac{1}{{2\nu + 1 \choose \nu + \frac{1}{2}}}\frac{{}_2 F_1\left(\nu - \frac{1}{2}, \nu + \frac{1}{2}, 2\nu + 1 ; -\xi^{-1}\right)}{\left(1 + \xi\right)\xi^{\nu + \frac{1}{2}}\left(x_3 y_3\right)}. \qquad
\end{IEEEeqnarray}
The $\mathfrak{su}\left(2\right)$ generators $t_i$ are expressed in terms of the fuzzy spherical harmonics $\hat{Y}^{m}_{\ell}$ as follows:
\begin{IEEEeqnarray}{l}
t_1 = \frac{\left(-1\right)^{k+1}}{2}\sqrt{\frac{k\left(k^2 - 1\right)}{6}} \cdot \left(\hat{Y}^{-1}_{1} - \hat{Y}^{1}_{1}\right) \label{t1FuzzySphericalHarmonics} \\[6pt]
t_2 = \frac{i\left(-1\right)^{k+1}}{2}\sqrt{\frac{k\left(k^2 - 1\right)}{6}} \cdot \left(\hat{Y}^{-1}_{1} + \hat{Y}^{1}_{1}\right) \label{t2FuzzySphericalHarmonics} \\[6pt]
t_3 = \frac{\left(-1\right)^{k+1}}{2}\sqrt{\frac{k\left(k^2 - 1\right)}{3}} \cdot \hat{Y}^{0}_{1}, \label{t3FuzzySphericalHarmonics}
\end{IEEEeqnarray}
so that by using \eqref{t1FuzzySphericalHarmonics}--\eqref{t3FuzzySphericalHarmonics} and the orthonormality relation
\begin{IEEEeqnarray}{l}
\text{tr}\left[\hat{Y}^{m}_{\ell} \hat{Y}^{m'}_{\ell'}\right] = \left(-1\right)^{m'} \delta_{\ell\ell'} \delta_{m+m'},
\end{IEEEeqnarray}
we arrive at the following color factors:
\begin{IEEEeqnarray}{l}
\sum_{i = 1}^{3} \sum_{m = -\ell}^{\ell} \left(-1\right)^{m}\text{tr}\left[t_i^{(k)} \cdot \hat{Y}^{m}_{\ell}\right] \text{tr}\left[t_i^{(k)} \cdot \hat{Y}^{-m}_{\ell}\right] = \frac{k\left(k^2 - 1\right)\delta_{\ell1}}{4} \qquad \label{ColorFactor1} \\[6pt]
i \sum_{i,j,l = 1}^{3} \sum_{m,m' = -\ell}^{\ell}\left(-1\right)^{m'} \epsilon_{ijl} \, \text{tr}\left[t_i^{(k)} \cdot \hat{Y}^{m}_{\ell}\right] \text{tr}\left[t_j^{(k)} \cdot \hat{Y}^{m'}_{\ell}\right] \times \nonumber \\[6pt]
\hspace{2cm} \times \left[t^{\left(2\ell + 1\right)}_{l}\right]_{\ell-m+1,\ell+m'+1} = \frac{k\left(k^2 - 1\right)\delta_{\ell1}}{2}. \qquad \label{ColorFactor2}
\end{IEEEeqnarray}

\bibliographystyle{elsarticle-num}
\bibliography{Bibliography}

\begin{thebibliography}{10}
\expandafter\ifx\csname url\endcsname\relax
  \def\url#1{\texttt{#1}}\fi
\expandafter\ifx\csname urlprefix\endcsname\relax\def\urlprefix{URL }\fi
\expandafter\ifx\csname href\endcsname\relax
  \def\href#1#2{#2} \def\path#1{#1}\fi

\bibitem{HerzogHuang17}
C.~P. Herzog, K.-W. Huang, {Boundary conformal field theory and a boundary
  central charge}, JHEP \textbf{10} (2017) 189.
\newblock \href {http://arxiv.org/abs/1707.06224} {\path{arXiv:1707.06224}},
  \href {https://doi.org/10.1007/JHEP10(2017)189}
  {\path{doi:10.1007/JHEP10(2017)189}}.

\bibitem{HerzogHuangJensen17}
C.~Herzog, K.-W. Huang, K.~Jensen, {Displacement operators and constraints on
  boundary central charges}, Phys. Rev. Lett. 120 (2018) 021601.
\newblock \href {http://arxiv.org/abs/1709.07431} {\path{arXiv:1709.07431}},
  \href {https://doi.org/10.1103/PhysRevLett.120.021601}
  {\path{doi:10.1103/PhysRevLett.120.021601}}.

\bibitem{KarchRandall01a}
A.~Karch, L.~Randall, {Localized gravity in string theory}, Phys. Rev. Lett.
  \textbf{87} (2001) 061601.
\newblock \href {http://arxiv.org/abs/hep-th/0105108}
  {\path{arXiv:hep-th/0105108}}, \href
  {https://doi.org/10.1103/PhysRevLett.87.061601}
  {\path{doi:10.1103/PhysRevLett.87.061601}}.

\bibitem{KarchRandall01b}
A.~Karch, L.~Randall, {Open and closed string interpretation of susy CFT's on
  branes with boundaries}, JHEP \textbf{06} (2001) 063.
\newblock \href {http://arxiv.org/abs/hep-th/0105132}
  {\path{arXiv:hep-th/0105132}}, \href
  {https://doi.org/10.1088/1126-6708/2001/06/063}
  {\path{doi:10.1088/1126-6708/2001/06/063}}.

\bibitem{ConstableMyersTafjord99}
N.~R. Constable, R.~C. Myers, O.~Tafjord, {The noncommutative bion core}, Phys.
  Rev. \textbf{D61} (2000) 106009.
\newblock \href {http://arxiv.org/abs/hep-th/9911136}
  {\path{arXiv:hep-th/9911136}}, \href
  {https://doi.org/10.1103/PhysRevD.61.106009}
  {\path{doi:10.1103/PhysRevD.61.106009}}.

\bibitem{MyersWapler08}
R.~C. Myers, M.~C. Wapler, {Transport properties of holographic defects}, JHEP
  \textbf{12} (2008) 115.
\newblock \href {http://arxiv.org/abs/0811.0480} {\path{arXiv:0811.0480}},
  \href {https://doi.org/10.1088/1126-6708/2008/12/115}
  {\path{doi:10.1088/1126-6708/2008/12/115}}.

\bibitem{KristjansenSemenoffYoung12b}
C.~Kristjansen, G.~W. Semenoff, D.~Young, {Chiral primary one-point functions
  in the D3-D7 defect conformal field theory}, JHEP \textbf{01} (2013) 117.
\newblock \href {http://arxiv.org/abs/1210.7015} {\path{arXiv:1210.7015}},
  \href {https://doi.org/10.1007/JHEP01(2013)117}
  {\path{doi:10.1007/JHEP01(2013)117}}.

\bibitem{Nahm79c}
W.~Nahm, {A simple formalism for the BPS monopole}, Phys. Lett. \textbf{B90}
  (1980) 413.
\newblock \href {https://doi.org/10.1016/0370-2693(80)90961-2}
  {\path{doi:10.1016/0370-2693(80)90961-2}}.

\bibitem{GaiottoWitten08a}
D.~Gaiotto, E.~Witten, {Supersymmetric boundary conditions in $\mathcal{N} = 4$
  super Yang-Mills theory}, J. Stat. Phys. \textbf{135} (2009) 789.
\newblock \href {http://arxiv.org/abs/0804.2902} {\path{arXiv:0804.2902}},
  \href {https://doi.org/10.1007/s10955-009-9687-3}
  {\path{doi:10.1007/s10955-009-9687-3}}.

\bibitem{deLeeuwKristjansenZarembo15}
M.~de~Leeuw, C.~Kristjansen, K.~Zarembo, {One-point functions in defect CFT and
  integrability}, JHEP \textbf{08} (2015) 098.
\newblock \href {http://arxiv.org/abs/1506.06958} {\path{arXiv:1506.06958}},
  \href {https://doi.org/10.1007/JHEP08(2015)098}
  {\path{doi:10.1007/JHEP08(2015)098}}.

\bibitem{Buhl-MortensenLeeuwKristjansenZarembo15}
I.~Buhl-Mortensen, M.~de~Leeuw, C.~Kristjansen, K.~Zarembo, {One-point
  functions in AdS/dCFT from matrix product states}, JHEP \textbf{02} (2016)
  052.
\newblock \href {http://arxiv.org/abs/1512.02532} {\path{arXiv:1512.02532}},
  \href {https://doi.org/10.1007/JHEP02(2016)052}
  {\path{doi:10.1007/JHEP02(2016)052}}.

\bibitem{KristjansenMullerZarembo20a}
C.~Kristjansen, D.~{M\"{u}ller}, K.~Zarembo, {Integrable boundary states in
  D3-D5 dCFT: beyond scalars}, JHEP \textbf{08} (2020) 103.
\newblock \href {http://arxiv.org/abs/2005.01392} {\path{arXiv:2005.01392}},
  \href {https://doi.org/10.1007/JHEP08(2020)103}
  {\path{doi:10.1007/JHEP08(2020)103}}.

\bibitem{Buhl-MortensenLeeuwIpsenKristjansenWilhelm17a}
I.~Buhl-Mortensen, M.~de~Leeuw, A.~Ipsen, C.~Kristjansen, M.~Wilhelm,
  {Asymptotic one-point functions in gauge-string duality with defects}, Phys.
  Rev. Lett. \textbf{119} (2017) 261604.
\newblock \href {http://arxiv.org/abs/1704.07386} {\path{arXiv:1704.07386}},
  \href {https://doi.org/10.1103/PhysRevLett.119.261604}
  {\path{doi:10.1103/PhysRevLett.119.261604}}.

\bibitem{GomborBajnok20a}
T.~Gombor, Z.~Bajnok, {Boundary states, overlaps, nesting and bootstrapping
  AdS/dCFT}, JHEP \textbf{10} (2020) 123.
\newblock \href {http://arxiv.org/abs/2004.11329} {\path{arXiv:2004.11329}},
  \href {https://doi.org/10.1007/JHEP10(2020)123}
  {\path{doi:10.1007/JHEP10(2020)123}}.

\bibitem{GomborBajnok20b}
T.~Gombor, Z.~Bajnok, {Boundary state bootstrap and asymptotic overlaps in
  AdS/dCFT}, JHEP 03 (2021) 222.
\newblock \href {http://arxiv.org/abs/2006.16151} {\path{arXiv:2006.16151}},
  \href {https://doi.org/10.1007/JHEP03(2021)222}
  {\path{doi:10.1007/JHEP03(2021)222}}.

\bibitem{KomatsuWang20}
S.~Komatsu, Y.~Wang, {Non-perturbative defect one-point functions in planar
  $\mathcal{N}=4$ super-Yang-Mills}, Nucl. Phys. \textbf{B958} (2020) 115120.
\newblock \href {http://arxiv.org/abs/2004.09514} {\path{arXiv:2004.09514}},
  \href {https://doi.org/10.1016/j.nuclphysb.2020.115120}
  {\path{doi:10.1016/j.nuclphysb.2020.115120}}.

\bibitem{deLeeuwGomborKristjansenLinardopoulosPozsgay19}
M.~de~Leeuw, T.~Gombor, C.~Kristjansen, G.~Linardopoulos, B.~Pozsgay, {Spin
  chain overlaps and the twisted Yangian}, JHEP \textbf{01} (2020) 176.
\newblock \href {http://arxiv.org/abs/1912.09338} {\path{arXiv:1912.09338}},
  \href {https://doi.org/10.1007/JHEP01(2020)176}
  {\path{doi:10.1007/JHEP01(2020)176}}.

\bibitem{deLeeuwKristjansenVardinghus19}
M.~de~Leeuw, C.~Kristjansen, K.~E. Vardinghus, {A non-integrable quench from
  AdS/dCFT}, Phys. Lett. \textbf{B798} (2019) 134940.
\newblock \href {http://arxiv.org/abs/1906.10714} {\path{arXiv:1906.10714}},
  \href {https://doi.org/10.1016/j.physletb.2019.134940}
  {\path{doi:10.1016/j.physletb.2019.134940}}.

\bibitem{DekelOz11b}
A.~Dekel, Y.~Oz, {Integrability of Green-Schwarz sigma models with boundaries},
  JHEP \textbf{08} (2011) 004.
\newblock \href {http://arxiv.org/abs/1106.3446} {\path{arXiv:1106.3446}},
  \href {https://doi.org/10.1007/JHEP08(2011)004}
  {\path{doi:10.1007/JHEP08(2011)004}}.

\bibitem{LinardopoulosZarembo21}
G.~Linardopoulos, K.~Zarembo, {String integrability of defect CFT and dynamical
  reflection matrices}, JHEP \textbf{05} (2021) 203.
\newblock \href {http://arxiv.org/abs/2102.12381} {\path{arXiv:2102.12381}},
  \href {https://doi.org/10.1007/JHEP05(2021)203}
  {\path{doi:10.1007/JHEP05(2021)203}}.

\bibitem{deLeeuwIpsenKristjansenWilhelm17}
M.~de~Leeuw, A.~C. Ipsen, C.~Kristjansen, M.~Wilhelm, {Introduction to
  integrability and one-point functions in $\mathcal{N} = 4$ SYM and its defect
  cousin}, Les Houches Lect.\ Notes 106 (2019).
\newblock \href {http://arxiv.org/abs/1708.02525} {\path{arXiv:1708.02525}},
  \href {https://doi.org/10.1093/oso/9780198828150.003.0008}
  {\path{doi:10.1093/oso/9780198828150.003.0008}}.

\bibitem{deLeeuw19}
M.~de~Leeuw, {One-point functions in AdS/dCFT}, J. Phys. \textbf{A53} (2020)
  283001.
\newblock \href {http://arxiv.org/abs/1908.03444} {\path{arXiv:1908.03444}},
  \href {https://doi.org/10.1088/1751-8121/ab15fb}
  {\path{doi:10.1088/1751-8121/ab15fb}}.

\bibitem{Linardopoulos20}
G.~Linardopoulos, {Solving holographic defects}, PoS Corfu2019 (2020) 141.
\newblock \href {http://arxiv.org/abs/2005.02117} {\path{arXiv:2005.02117}}.

\bibitem{BianchiLemos19}
L.~Bianchi, M.~Lemos, {Superconformal surfaces in four dimensions}, JHEP
  \textbf{06} (2020) 056.
\newblock \href {http://arxiv.org/abs/1911.05082} {\path{arXiv:1911.05082}},
  \href {https://doi.org/10.1007/JHEP06(2020)056}
  {\path{doi:10.1007/JHEP06(2020)056}}.

\bibitem{McAvityOsborn95}
D.~M. McAvity, H.~Osborn, {Conformal field theories near a boundary in general
  dimensions}, Nucl. Phys. \textbf{B455} (1995) 522.
\newblock \href {http://arxiv.org/abs/cond-mat/9505127}
  {\path{arXiv:cond-mat/9505127}}, \href
  {https://doi.org/10.1016/0550-3213(95)00476-9}
  {\path{doi:10.1016/0550-3213(95)00476-9}}.

\bibitem{LiendoRastellivanRees12}
P.~Liendo, L.~Rastelli, B.~C. van Rees, {The bootstrap program for boundary
  CFT$_d$}, JHEP \textbf{07} (2013) 113.
\newblock \href {http://arxiv.org/abs/1210.4258} {\path{arXiv:1210.4258}},
  \href {https://doi.org/10.1007/JHEP07(2013)113}
  {\path{doi:10.1007/JHEP07(2013)113}}.

\bibitem{McAvityOsborn93}
D.~M. McAvity, H.~Osborn, {Energy momentum tensor in conformal field theories
  near a boundary}, Nucl. Phys. \textbf{B406} (1993) 655.
\newblock \href {http://arxiv.org/abs/hep-th/9302068}
  {\path{arXiv:hep-th/9302068}}, \href
  {https://doi.org/10.1016/0550-3213(93)90005-A}
  {\path{doi:10.1016/0550-3213(93)90005-A}}.

\bibitem{Fursaev15}
D.~Fursaev, {Conformal anomalies of CFT's with boundaries}, JHEP 12 (2015) 112.
\newblock \href {http://arxiv.org/abs/1510.01427} {\path{arXiv:1510.01427}},
  \href {https://doi.org/10.1007/JHEP12(2015)112}
  {\path{doi:10.1007/JHEP12(2015)112}}.

\bibitem{Buhl-MortensenLeeuwIpsenKristjansenWilhelm16c}
I.~Buhl-Mortensen, M.~de~Leeuw, A.~C. Ipsen, C.~Kristjansen, M.~Wilhelm, {A
  quantum check of AdS/dCFT}, JHEP \textbf{01} (2017) 098.
\newblock \href {http://arxiv.org/abs/1611.04603} {\path{arXiv:1611.04603}},
  \href {https://doi.org/10.1007/JHEP01(2017)098}
  {\path{doi:10.1007/JHEP01(2017)098}}.

\bibitem{BirrellDavies99}
N.~D. Birrell, P.~C.~W. Davies, {Quantum fields in curved space}, Cambridge
  University Press, Cambridge, 1999.

\bibitem{CallanColemanJackiw70}
C.~G. Callan~Jr., S.~R. Coleman, R.~Jackiw, {A new improved energy-momentum
  tensor}, Annals Phys. \textbf{59} (1970) 42.
\newblock \href {https://doi.org/10.1016/0003-4916(70)90394-5}
  {\path{doi:10.1016/0003-4916(70)90394-5}}.

\bibitem{OsbornPetkou93}
H.~Osborn, A.~C. Petkou, {Implications of conformal invariance in field
  theories for general dimensions}, Annals Phys. \textbf{231} (1994) 311.
\newblock \href {http://arxiv.org/abs/hep-th/9307010}
  {\path{arXiv:hep-th/9307010}}, \href {https://doi.org/10.1006/aphy.1994.1045}
  {\path{doi:10.1006/aphy.1994.1045}}.

\bibitem{Osborn19}
H.~Osborn, \href{https://www.damtp.cam.ac.uk/user/ho/CFTNotes.pdf}{Lectures on
  conformal field theories in more than two dimensions} (2019).

\bibitem{ErdmengerOsborn96}
J.~Erdmenger, H.~Osborn, {Conserved currents and the energy momentum tensor in
  conformally invariant theories for general dimensions}, Nucl. Phys.
  \textbf{B483} (1997) 431.
\newblock \href {http://arxiv.org/abs/hep-th/9605009}
  {\path{arXiv:hep-th/9605009}}, \href
  {https://doi.org/10.1016/S0550-3213(96)00545-7}
  {\path{doi:10.1016/S0550-3213(96)00545-7}}.

\bibitem{HoweSokatchevWest98}
P.~S. Howe, E.~Sokatchev, P.~C. West, {Three point functions in $\mathcal{N} =
  4$ Yang-Mills}, Phys. Lett. \textbf{B444} (1998) 341.
\newblock \href {http://arxiv.org/abs/hep-th/9808162}
  {\path{arXiv:hep-th/9808162}}, \href
  {https://doi.org/10.1016/S0370-2693(98)01431-2}
  {\path{doi:10.1016/S0370-2693(98)01431-2}}.

\bibitem{Buhl-MortensenLeeuwIpsenKristjansenWilhelm16a}
I.~Buhl-Mortensen, M.~de~Leeuw, A.~C. Ipsen, C.~Kristjansen, M.~Wilhelm,
  {One-loop one-point functions in gauge-gravity dualities with defects}, Phys.
  Rev. Lett. \textbf{117} (2016) 231603.
\newblock \href {http://arxiv.org/abs/1606.01886} {\path{arXiv:1606.01886}},
  \href {https://doi.org/10.1103/PhysRevLett.117.231603}
  {\path{doi:10.1103/PhysRevLett.117.231603}}.

\bibitem{deLeeuwIpsenKristjansenVardinghusWilhelm17}
M.~de~Leeuw, A.~C. Ipsen, C.~Kristjansen, K.~E. Vardinghus, M.~Wilhelm,
  {Two-point functions in AdS/dCFT and the boundary conformal bootstrap
  equations}, JHEP \textbf{08} (2017) 020.
\newblock \href {http://arxiv.org/abs/1705.03898} {\path{arXiv:1705.03898}},
  \href {https://doi.org/10.1007/JHEP08(2017)020}
  {\path{doi:10.1007/JHEP08(2017)020}}.

\bibitem{DeWolfeFreedmanOoguri01}
{DeWolfe, O.}, {Freedman, D.Z.}, {Ooguri, H.}, {Holography and defect conformal
  field theories}, Phys. Rev. \textbf{D66} (2002) 025009.
\newblock \href {http://arxiv.org/abs/hep-th/0111135}
  {\path{arXiv:hep-th/0111135}}, \href
  {https://doi.org/10.1103/PhysRevD.66.025009}
  {\path{doi:10.1103/PhysRevD.66.025009}}.

\bibitem{NagasakiYamaguchi12}
K.~Nagasaki, S.~Yamaguchi, {Expectation values of chiral primary operators in
  holographic interface CFT}, Phys. Rev. \textbf{D86} (2012) 086004.
\newblock \href {http://arxiv.org/abs/1205.1674} {\path{arXiv:1205.1674}},
  \href {https://doi.org/10.1103/PhysRevD.86.086004}
  {\path{doi:10.1103/PhysRevD.86.086004}}.

\bibitem{GeorgiouLinardopoulosZoakos23}
G.~Georgiou, G.~Linardopoulos, D.~Zoakos, {Holographic correlators of
  semiclassical states in defect CFTs} (2023).
\newblock \href {http://arxiv.org/abs/2304.10434} {\path{arXiv:2304.10434}}.

\bibitem{RastelliZhou17a}
L.~Rastelli, X.~Zhou, {The Mellin formalism for boundary CFT$_d$}, JHEP
  \textbf{10} (2017) 146.
\newblock \href {http://arxiv.org/abs/1705.05362} {\path{arXiv:1705.05362}},
  \href {https://doi.org/10.1007/JHEP10(2017)146}
  {\path{doi:10.1007/JHEP10(2017)146}}.

\bibitem{GoncalvesItsios18}
V.~Goncalves, G.~Itsios, {A note on defect Mellin amplitudes} (2018).
\newblock \href {http://arxiv.org/abs/1803.06721} {\path{arXiv:1803.06721}}.

\bibitem{KavirajPaulos18}
A.~Kaviraj, M.~F. Paulos, {The functional bootstrap for boundary CFT}, JHEP
  \textbf{04} (2020) 135.
\newblock \href {http://arxiv.org/abs/1812.04034} {\path{arXiv:1812.04034}},
  \href {https://doi.org/10.1007/JHEP04(2020)135}
  {\path{doi:10.1007/JHEP04(2020)135}}.

\bibitem{MazacRastelliZhou18}
D.~Maz\'a\v{c}, L.~Rastelli, X.~Zhou, {An analytic approach to BCFT$_{d}$},
  JHEP \textbf{12} (2019) 004.
\newblock \href {http://arxiv.org/abs/1812.09314} {\path{arXiv:1812.09314}},
  \href {https://doi.org/10.1007/JHEP12(2019)004}
  {\path{doi:10.1007/JHEP12(2019)004}}.

\bibitem{KastikainenShashi21}
J.~Kastikainen, S.~Shashi, {Structure of holographic BCFT correlators from
  geodesics}, Phys. Rev. \textbf{D105} (2022) 046007.
\newblock \href {http://arxiv.org/abs/2109.00079} {\path{arXiv:2109.00079}},
  \href {https://doi.org/10.1103/PhysRevD.105.046007}
  {\path{doi:10.1103/PhysRevD.105.046007}}.

\bibitem{ChenZhou23}
J.~Chen, X.~Zhou, {Aspects of higher-point functions in BCFT$_d$} (2023).
\newblock \href {http://arxiv.org/abs/2304.11799} {\path{arXiv:2304.11799}}.

\bibitem{GimenezGrau23}
A.~Gimenez-Grau, {The Witten diagram bootstrap for holographic defects} (2023).
\newblock \href {http://arxiv.org/abs/2306.11896} {\path{arXiv:2306.11896}}.

\end{thebibliography}
\end{document}